\documentclass[global,twocolumn]{svjour}
\usepackage{graphics}
\usepackage{isotope}
\journalname{Applied Physics B}
\begin{document}
\title{Inserting single Cs atoms into an ultracold Rb gas}
\author{Nicolas Spethmann \inst{1}, Farina Kindermann \inst{1}, Shincy John \inst{1}, Claudia Weber \inst{1}, Dieter Meschede \inst{1} \and Artur Widera\inst{2}
}                     
\offprints{}          
\institute{Institut f\"ur Angewandte Physik, Universit\"at Bonn, 53115 Bonn, Germany \and Fachbereich Physik und Landesforschungszentrum OPTIMAS, Universit\"at Kaiserslautern, 67663 Kaiserslautern, Germany }
\date{Received: date / Revised version: date}
%
\maketitle
\begin{abstract}
We report on the controlled insertion of individual Cs atoms into an ultracold Rb gas at $\approx$ 400 nK. This requires to combine the techniques necessary for cooling, trapping and manipulating single laser cooled atoms around the Doppler temperature with an experiment to produce ultracold degenerate quantum gases. In our approach, both systems are prepared in separated traps and then combined. Our results pave the way for coherent interaction between a quantum gas and a single or few neutral atoms of another species. 
\end{abstract}
\section{Introduction}
\label{intro}

Ultracold atoms form a versatile model system to study various regimes of quantum phenomena \cite{bloch_many-body_2008}. Owing to advances in control of single particles, even the controlled interaction between single individual impurity atoms and a quantum gas seems to become feasible, which could shed light onto questions such as impurity physics and polaron formation in quantum gases \cite{cucchietti_strong-coupling_2006,kalas_interaction-induced_2006}, or novel coherent cooling mechanisms \cite{griessner_dark-state_2006}. Recently, single atom resolution in single species quantum gases has been achieved \cite{bakr_quantum_2009,sherson_single-atom-resolved_2010,gericke_high-resolution_2008}, and the immersion of a single ion in a Bose-Einstein condensate (BEC) has been realized \cite{zipkes_trapped_2010,schmid_dynamics_2010}. Our work aims at the insertion of single or few neutral  Cs atoms into a Rb BEC.


Typical single atom experiments work with laser cooled atoms that are loaded from a magneto-optical trap (MOT) and thus have temperatures in the range of $\approx$ 100 $\mu$K. These experiments feature a high level of control over all degrees of freedom \cite{foerster_microwave_2009}, and single atom resolution is intrinsically given \cite{PhysRevLett.102.053001}. However, the density is very low and it is challenging to introduce coherent interaction between several individual atoms. Experiments dealing with quantum gases prepared by evaporative cooling, in contrast, provide samples at ultralow temperatures ($\approx$ 100 nK) and very high densities (10$^{13}$-10$^{15}$ / cm$^3$). In these systems, coherent interactions dominate. Single atom control and resolution, however, are quite challenging and usually bulk properties of the ensemble are investigated. The idea of our work is to combine the advantages of single atom experiments with the benefits of quantum gases. 

In our approach, we first produce an ultracold Rb gas, and move it close to the position of a high gradient single Cs atom's MOT. The Rb gas is transferred into an optical trap during the transport. Then individual atoms are captured in the high gradient MOT. The spatial overlap between the two species is measured employing interspecies light-induced collisions, allowing for precise adjustment of their relative positions. This facilitates trapping both species in separated sites of an optical lattice and subsequent transfer of both, Cs atoms and ultracold Rb gas, into the same purely optical potential.

The paper is structured as follows: We will first briefly review the properties of our QUIC trap and describe the magnetic transport of the ultracold Rb cloud. Then we discuss the addition of an optical trap to provide constant and relatively large trap frequencies during the transport. Subsequently, we introduce the high gradient MOT which is used to trap single Cs atoms. An optical lattice allows trapping of both species at separated lattice sites. Finally, we demonstrate the optimization of inserting single Cs atoms into the Rb gas.

\section{Production and transport of the Rb gas}
\label{transport}

\begin{figure}
    \resizebox{0.48\textwidth}{!}{%
    \includegraphics{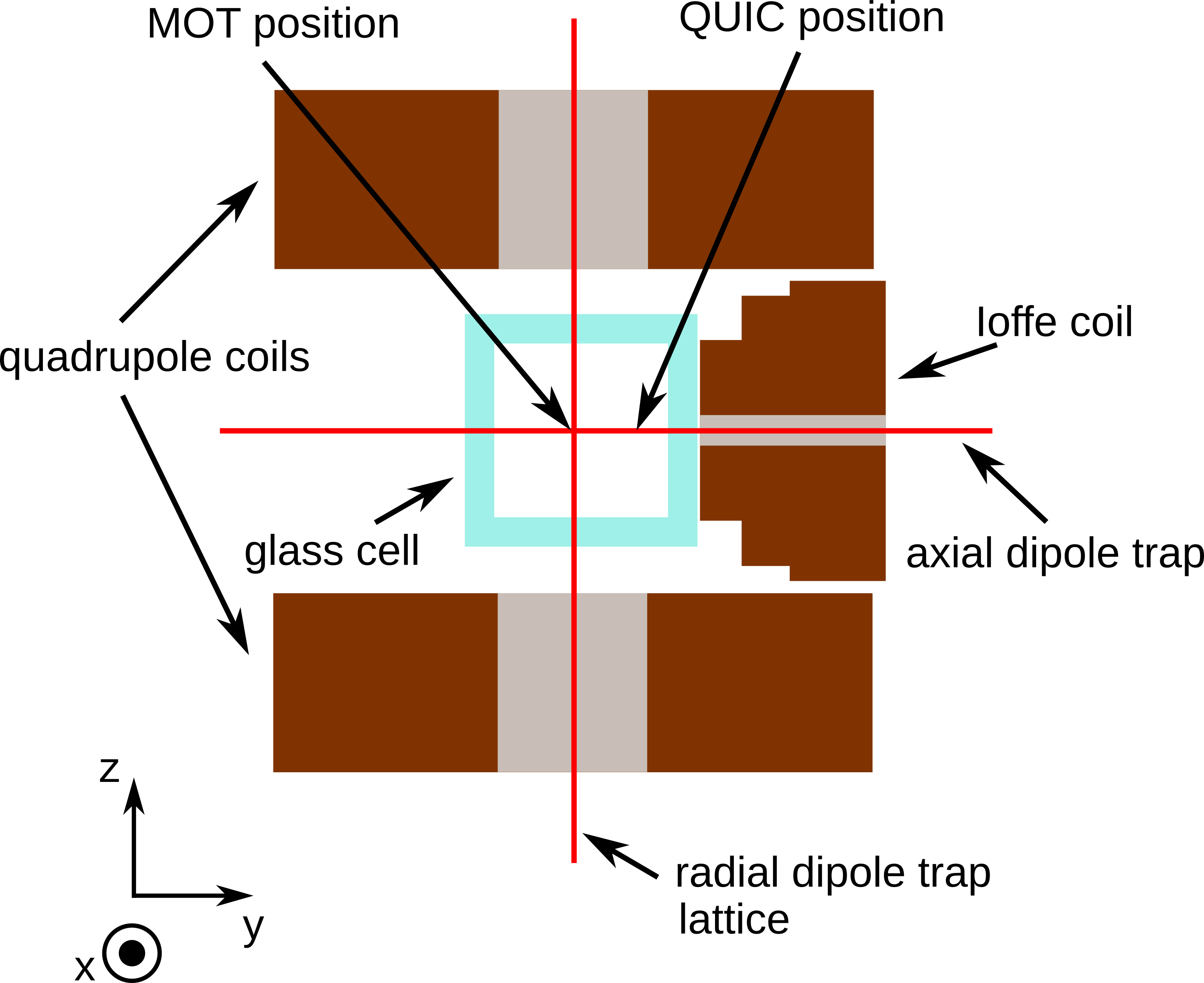}
    }  
  \caption{Schematic setup of the experiment. Single Cs atoms can be trapped in a high gradient MOT at the center ($x=y=z=0$) of the apparatus. Ultracold Rb clouds are prepared in the QUIC-type magnetic trap at a position  $\approx$ 7 mm apart from the center ($y\approx$ 7 mm$, x=z=0$). By the guided magnetic transport, the Rb gas is moved along the $y$-direction to the center in close vicinity to the single atom MOT.}
\label{setup}
\end{figure}

\begin{figure*}
 \resizebox{1.0\textwidth}{!}{%
    \includegraphics{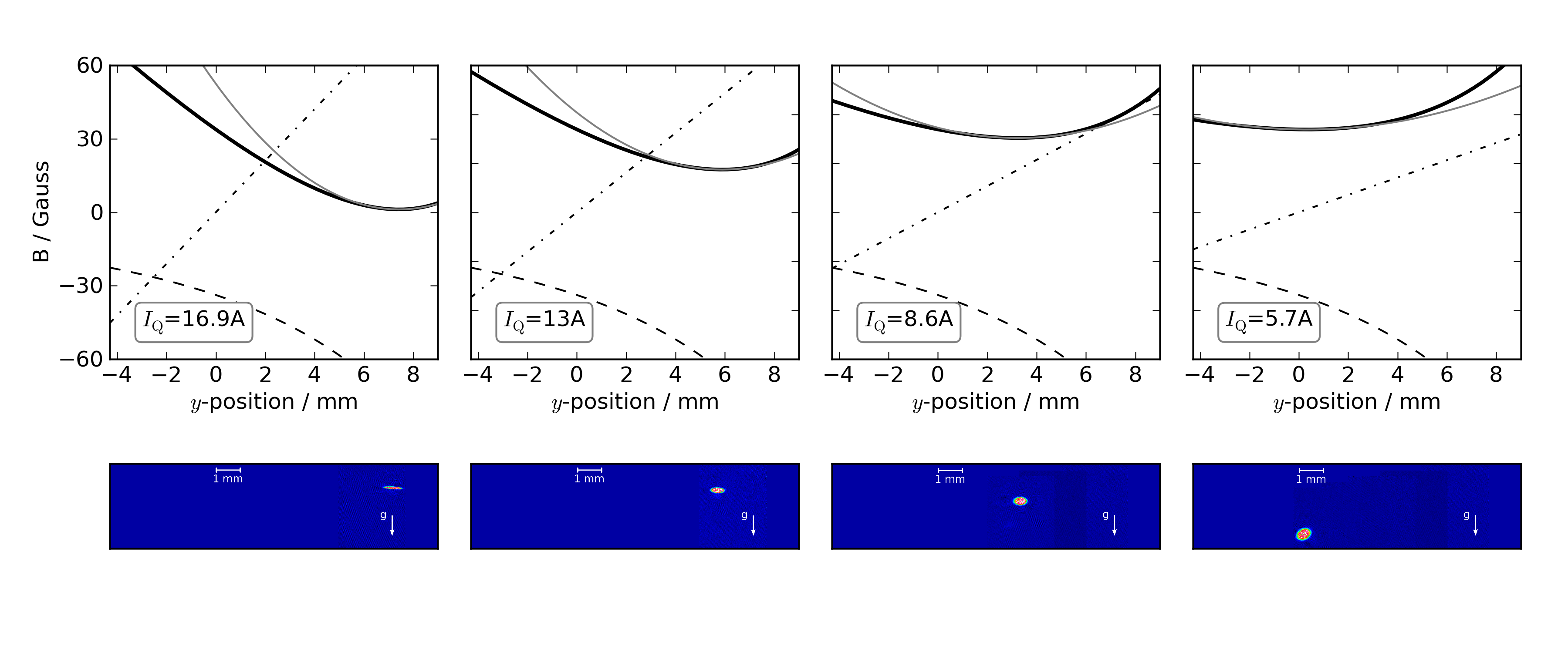}
    }
\caption{Purely magnetic transport at four different stages. From left to right: $I_\mathrm{Q}=16.9$ A, $I_\mathrm{Q}=13$ A, $I_\mathrm{Q}=8.6$ A, $I_\mathrm{Q}=5.7$ A. \textbf{Upper row:} Cut through the magnetic field in $y$-direction at different transport stages. Shown are the Ioffe field (dashed line), the quadrupole field (dashed-dotted line) and the absolute value of the effective magnetic field $|\vec{B}_\mathrm{QUIC}|$, along with the harmonic approximation (light gray). \textbf{Lower row}: Absorption images of the Rb cloud in the magnetic trap corresponding to the field configuration shown in the upper panel. Due to gravitational sag $g/\omega_\mathrm{r}^2$, the center of the trap shifts also in $x$-direction (see also Fig. \ref{pos_omega_vs_I}b). To cover the entire field of view of more than 7mm, for this measurement the CCD camera was moved laterally several times. The position of each image was calibrated such that it can be displayed in a sufficiently large frame. The remaining part of the frame was filled with the color corresponding to the background.}
\label{transport_overview}
\end{figure*}

In our experiment BECs in the $|2,2\hspace{-0.2cm}>$-state are produced in a QUIC-type magnetic trap \cite{esslinger_bose-einstein_1998}. For details of the apparatus see \cite {haas_species-selective_2007}. In this type of trap, a Ioffe and quadrupole field nearly cancel at the trap center, which is in our case at a distance of about 7\,mm from the zero-crossing of the quadrupole field (see Figs. \ref{setup} and \ref{transport_overview}). The magnetic field can be approximated by a combination of a quadrupole and a dipole field
\begin{equation}
 \vec{B}_{\mathrm{QUIC}}(\vec{r}) = \vec{B}_\mathrm{Q}(\vec{r}) + \vec{B}_\mathrm{I}(\vec{r}),	
\label{QUIC}
\end{equation}
where
\begin{equation}
  \vec{B}_\mathrm{Q}(\vec{r}) = \xi(\vec{e}_{x} + \vec{e}_{y} - 2\vec{e}_{z})
\end{equation}
and
\begin{equation}
  \vec{B}_\mathrm{I}(\vec{r}'=\vec{r} - y_\mathrm{I}\vec{e}_{y}) = \frac{3py'\vec{r}'-p{r'}^2\vec{e}_{y}}{{r'}^5}.
\end{equation}

Here, $\xi \approx 6.2 \,$ G/(A cm) is the gradient of the quad-rupole coils, $p \approx 27.4 \,$ G cm$^3/ \textnormal{A}$  is the dipole moment of the Ioffe coil and $y_\mathrm{I}\approx 30$ mm is the distance between Ioffe and quadrupole coils.

The  remaining offset field at the center of the trap is adjusted to $B_\mathrm{0}\approx 1 \mathrm{G}$, in order to avoid Majorana losses due to spin flips. By exciting the center of mass motion of trapped atoms we measured the trap frequencies in this compressed trap (at $I_\mathrm{Q}=I_\mathrm{I}=16.9$ A) to be $\omega_\mathrm{r} = 2 \pi \times 179$ Hz ($\omega_\mathrm{a} = 2 \pi \times 17.9$ Hz) for the radial direction along $x$ and $z$ (axial direction along $y$), in good agreement with calculated values using $\xi$ and $p$.

By decreasing the quadrupole field strength, the center of the trap moves towards the quadrupole field zero crossing at the center of the apparatus ($x=y=z=0$). Using (\ref{QUIC}), the position of the trap center $y_0$ can be calculated to be

\begin{equation}
 y_\mathrm{0} = -\left(\frac{6 I_\mathrm{I}p}{I_\mathrm{Q}\xi}\right)^{\frac{1}{4}}+y_\mathrm{I}.
  \label{y0}
\end{equation}

Here, $I_\mathrm{Q}$ is the current running through the quadrupole coil and $I_\mathrm{I}$ is the current of the Ioffe coil. In the top row of Fig. \ref{transport_overview}, the corresponding magnetic fields along the transport direction ($y$-direction, $x=z=0$) are plotted, showing how the field components add up to the effective field for varying values of $I_Q$. In the bottom row of Fig. \ref{transport_overview} absorption images of a few million atoms with a temperature of $\approx$ 1 $\mu$K in the corresponding trap are presented. These images illustrate that, upon decreasing $I_Q$, the cloud moves towards the center of the apparatus ($y$=0). At the same time, however, the confinement decreases, leading to a larger thermal cloud and a larger gravitational sag. 

Both observations, changing position and confinement, are quantified in Fig. \ref{pos_omega_vs_I}. The atoms move along the $y$-direction as the quadrupole current $I_\mathrm{Q}$ is decreased (Fig. \ref{pos_omega_vs_I}a), in good agreement with the expectation according to (\ref{y0}). It should be noted that the new trap center is given by the point where the curvature of the quadrupole field starts to be bigger than the curvature of the Ioffe field rather than its absolute value. Therefore it is also possible to move the trap beyond the zero crossing of the quadrupole field. 

Fig. \ref{pos_omega_vs_I}b shows the measured radial trap frequency $\omega_\mathrm{r}$ (along the $x$-direction) during the transport. We model this decay of $\omega_\mathrm{r}$ as a function of quadrupole current $I_\mathrm{Q}$. This is important to counteract the relaxing confinement by an optical guiding field, discussed in section \ref{DT}. Knowing the expression for the magnetic field (\ref{QUIC}), $\omega_\mathrm{r}$ can be calculated by a harmonic approximation in the center of the trap
\begin{equation}
  B_{x\mathrm{,harmonic}} = B_\mathrm{0} +  \frac{\partial^2|\vec{B}_\mathrm{QUIC}|}{\partial x^2}x^2,
  \label{second_order_approx}
\end{equation}
where
\begin{eqnarray}
	  &\frac{\partial^2|\vec{B}_\mathrm{QUIC}|}{\partial x^2}  =  & \nonumber \\
 &\frac{1}{2}\left(\frac{(I_\mathrm{Q}\xi y^4 - 3I_\mathrm{Q}\xi y^3y_\mathrm{I} + 3I_\mathrm{Q}\xi y^2y_\mathrm{I}^2 - I_\mathrm{Q}\xi yy_\mathrm{I}^3 + 2I_\mathrm{I}p)^2}{(y_\mathrm{I}-y)^6}  \right)^\frac{1}{2} & \nonumber \\
 		&\times	\frac{(I_\mathrm{Q}\xi(y_\mathrm{I}-y)^5-3I_\mathrm{I}py + 3I_\mathrm{I}py_\mathrm{I})^2}{(y_\mathrm{I}-y)^4(I_\mathrm{Q}\xi y^4 - 3I_\mathrm{Q}\xi y^3y_\mathrm{I} + 3I_\mathrm{Q}\xi y^2y_\mathrm{I}^2-I_\mathrm{Q}\xi yy_\mathrm{I}^3+2I_\mathrm{I}p)^2}& \nonumber
  \label{d2Bdx}
\end{eqnarray}
and
\begin{equation}
  	  B_\mathrm{0} = |\vec{B}_\mathrm{QUIC}(y_\mathrm{0})|. \nonumber
\end{equation}
The radial trap frequency $\omega_\mathrm{r}$ is then given by
\begin{equation}
    V = \vec{\mu}_\mathrm{B}m_\mathrm{F}B_{x\mathrm{,harmonic}} = \frac{m_\mathrm{Rb}}{2}\omega_\mathrm{r}^2x^2.
\label{pot_theo}
\end{equation}

Here, $m_\mathrm{Rb}$ is the mass of the \isotope[87]{Rb} atom, and $\vec{\mu}_\mathrm{B}$ the Bohr magneton. The measured values for $\omega_\mathrm{r}$ presented in Fig. \ref{pos_omega_vs_I}b are in good agreement with our model according to (\ref{pot_theo}). 

It should be noted that (\ref{second_order_approx}) approximates the field on the symmetry axis of the QUIC coils ($x=z=0$). Due to gravitational sag the atoms are shifted away from this axis ($x\neq0$), which in turn leads to a change in the magnetic potential. This effect only becomes noticeable for very low trap frequencies, or long transport distances. The agreement with the experimental observation, however, is good for all parameters considered. Also, using the optical guiding field (see section \ref{DT}), the cloud is forced to stay approximately on the symmetry axis, and the knowledge of trap frequency decay at the start of the transport is sufficient.

The goal of the transport is a Rb sample with high atomic density at the Cs MOT position at $x=y=z=0$. In this regard, both expansion and sag ($\approx$ 2 mm) are not favorable. 

\section{Dipole trap as guiding field}
\label{DT}

The limitations of the magnetic transport discussed above can partially be avoided by using an additional far detuned optical dipole trap as guiding field. For the dipole trap, we use a Nd:YAG laser at 1064nm boosted by a fibre amplifier. The power is divided to provide two beams to eventually form a crossed dipole trap. One beam is propagating in the axial ($y$) direction through a hole in the Ioffe coil (see also Fig. \ref{setup}), forming the ''axial`` optical trap. The other beam propagates along the radial ($z$) direction, forming the ''radial`` trap. In Tab. \ref{table_parameter} the properties of the various traps are summarized. The power of each of the beams is stabilized by a home-built PID controller to a maximum power of 3 W (axially) or 0.6 W (radially). 

\begin{center}
\begin{table}
\begin{center}
\begin{tabular}{|l|c|c|c|c|}
  \hline
                           & QUIC   & radial & axial & lattice  \\
  \hline
  Rb depth/$k_B$ ($\mu$K)  &    $>$1000      &  27             & 30                &  67  \\
  Cs depth/$k_B$ ($\mu$K)  & -               &  49             & 54                &   720\\
  Wavelength (nm)          & -               &  1064           & 1064              & 899.9  \\    
  Waist/size ($\mu$m)      & (12,12,120)     &  48             & 100               &  31 \\    
  Rb rad. freq. (kHz)	   & \textbf{0.179}  &  \textbf{0.33}  & \textbf{0.17}     &  0.9 \\
  Rb ax. freq. (kHz)	   & \textbf{0.018}  &  0.002          & $<$0.001          &  138 \\
  Cs rad. freq. (kHz)	   & -               &  0.37           & 0.18              &   \textbf{2.1} \\        
  Cs ax. freq. (kHz)	   & -               &  0.002          & $<$0.001          & \textbf{311} \\ 
  \hline
\end{tabular} 
\caption{Parameter of the traps at maximum available laser intensity or magnetic field. Some parameters are measured (bold face), all others are calculated from known paramters.}
\label{table_parameter}
\end{center}
\end{table}
\end{center}

\begin{figure}
    \resizebox{0.5\textwidth}{!}{%
    \includegraphics{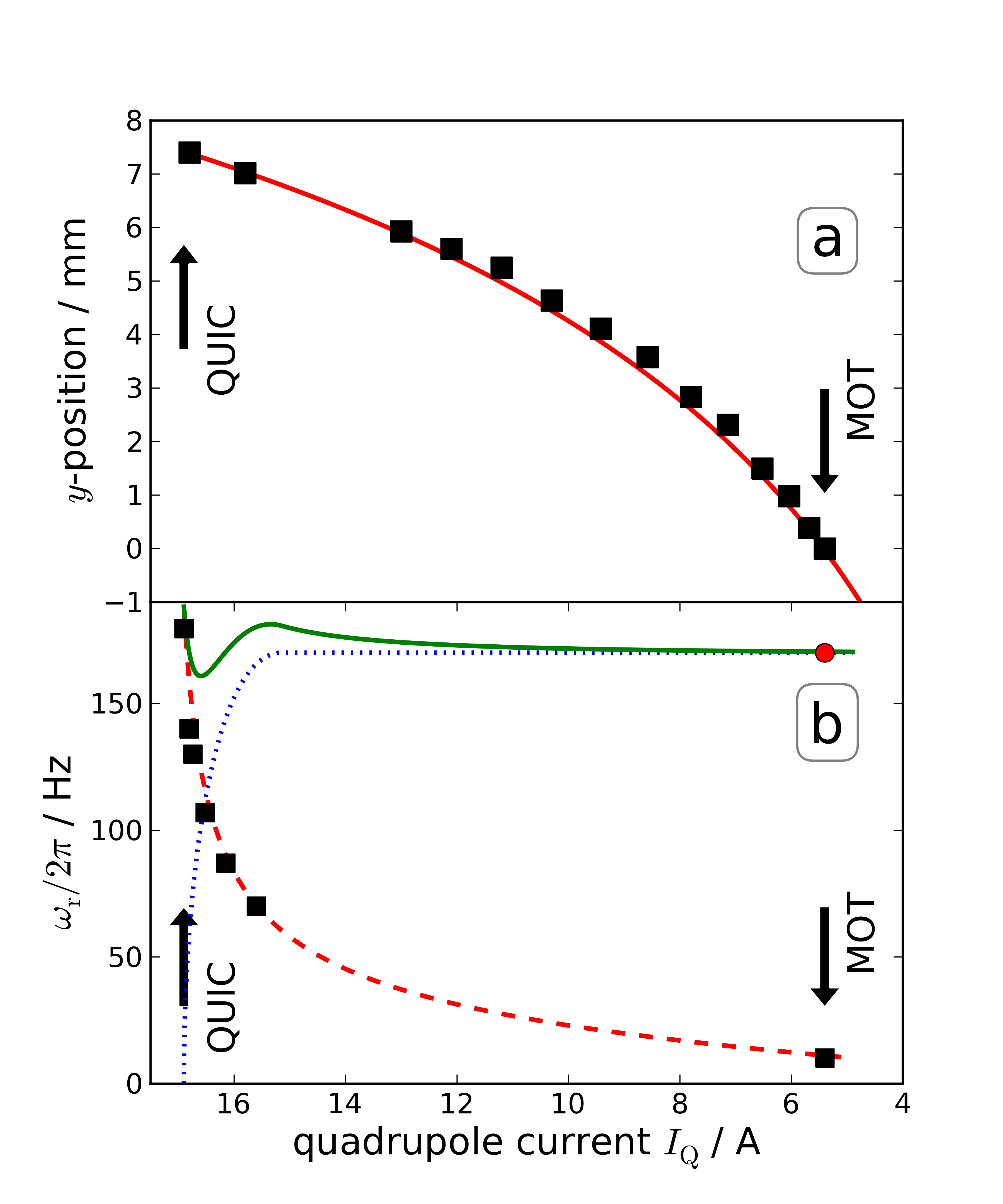}
    }
  \caption{\textbf{a)} Purely magnetic transport: By ramping down the quadrupole current $I_\mathrm{Q}$ while keeping the Ioffe current $I_\mathrm{I}$ constant, the ultracold gas can be moved from the original QUIC position (left side, $I_\mathrm{Q} = 16.9$ A) across a distance of about 7 mm to the center of the apparatus (Cs MOT position, $I_\mathrm{Q} \approx$ 5 A). The red solid line shows the expected position according to (\ref{y0}). \textbf{b)} Measured trap frequencies at different transport stages. The radial trap frequency in the QUIC trap of $\omega_\mathrm{r} = 2\pi \times 179$ Hz is strongly decreased to about $2\pi \times 10$ Hz at the end of the purely magnetic transport. The dashed red line shows the expected trap frequencies according to (\ref{pot_theo}). The decay of the trap frequencies is compensated for by using a dipole trap as guiding field (blue dotted line), see section \ref{DT}. The effective trap frequencies of the guided magnetic transport are approximately constant (green solid line). The errorbars for the presented data are of comparable size as the markersize and are therefore not shown.}
\label{pos_omega_vs_I}
\end{figure}

During the transport, the power for the axial dipole trap is ramped up, such that $\omega_\mathrm{r} = \sqrt{\omega_\mathrm{QUIC}^2 + \omega_\mathrm{dipole}^2}$ is kept approximately constant in the combined magnetic and optical potential (see Fig. \ref{pos_omega_vs_I}b), calculating $\omega_\mathrm{QUIC}$ using (\ref{pot_theo}). In this way, a ``mode matching'' of the cloud into the dipole trap is achieved. The magnetic transport then shifts the atomic cloud along the guiding dipole field. This gradual transfer also makes the transport and the loading into the dipole trap less sensitive to imperfections. For a reasonable range of positions of the axial dipole trap at the center of the apparatus, the tranport works without noticeable degradation in phase space density. This is important because later on this position of the axial dipole trap will be used to change the relative positions of Rb gas and single atom Cs MOT. 

To provide a reasonable confinement also along the third direction ($y$-direction) for the final experiments with the single atom Cs MOT, the radial dipole trap is ramped up within 200 ms while the Rb is stored in the combined magnetic and optical potential in the center of the apparatus at $x\approx y\approx z\approx 0$. Then, the magnetic fields of the QUIC are quickly switched off (within 1 ms), leaving the sample stored in the purely optical trap. Immediately after switching off the QUIC field, a homogeneous offset field is switched on to provide a quantization axis.


At the end of the transport, the Rb gas is typically cooled by evaporation in the crossed dipole trap. With this final evaporative cooling an optically trapped BEC with atom numbers in the range of 10$^4\ldots 10^5$ atoms can be produced in the center of the apparatus. The advantage of evaporative cooling in the dipole trap rather than transporting a BEC is a significantly higher stability. Small imperfections in the transport that lead to heating can be compensated by evaporative cooling after the transport. The high trapping frequencies provided by the crossed dipole trap support efficient cooling. Starting the transport with a Rb gas at about 1 $\mu$K in the QUIC trap and further evaporating after the transport was experimentally identified as the optimal strategy to produce a BEC after transport.

Thus, the guided transport provides a means to prepare ultracold Rb clouds in the crossed dipole trap at a position that can be chosen in a large parameter range. This paves the way for the controlled insertion of single Cs atoms into the Rb gas.

\section{Single Atoms: MOT and optical lattice}
\label{Cs_MOT}
Operating a MOT for Cs with a high magnetic field gradient produced by the quadrupole coils allows to cool and trap a small number (1-10) of Cs atoms. The exact number of atoms is inferred from a fluorescence measurement. For details of the high gradient Cs MOT see Ref. \cite{weber_single_2010}. 

To adjust the number of Cs atoms loaded, we operate the MOT for a time of typically 150 ms with a gradient around 60 G/cm. In this way we adjust the Poissoinian expectation value of the Cs atom number, typically to a value around two atoms. After this loading stage, the MOT is compressed by ramping up the gradient in a few ms to 300 G/cm, leading to a tighter MOT with a diameter of approximately 30 $\mu$m. In this high gradient MOT, the loading rate is typically below 1 atom/second, such that we keep the number of initially trapped atoms in the MOT. Cs-Cs light-induced collisions can be neglected due to the low Cs atom number \cite{ueberholz_b._cold_2002}.

In order to store Cs in a conserative potential, we add a one-dimensional optical lattice to the MOT. The lattice beams propagate along the $z$-direction, superimposed with the radial dipole trap. Due to the relative high temperature of Cs in the MOT ($\approx$ 100 $\mu$K), the potential has to be significantly larger than for Rb. Here, we focus 100 mW at 899.9 nm to a waist of 31 $\mu$m to create a trap depth of 720 $\mu$K. This is sufficient to transfer the Cs from the MOT into the lattice with an efficiency close to unity. At the same time Rb is loaded into the lattice, but due to the larger detuning for Rb, the potential is much more shallow aiding an adiabatic loading. In the lattice, the temperature of Cs was measured both with the adiabatic lowering technique \cite{alt_single_2003} and the release-recapture methode \cite{tuchendler_energy_2008} to be $\approx$ 27 $\mu$K.

At the end of each experiment Cs is recaptured in the high gradient MOT and the number of remaining Cs atoms is counted.

\section{Inserting single Cs atoms into the Rb gas}
\label{overlap}
\subsection{Species-selective trapping}

To cool and trap two different atomic species in close proximity, a species-selective trap is necessary, for example by the choice of the dipole trap wavelength \cite{leblanc_species-specific_2007}. In our case the selectivity is created by operating a high-gradient MOT for Cs while Rb is optically trapped in a $m_F=0$ Zeeman sublevel. Due to the large difference in wavelengths of the D2-transitions of Rb and Cs (780 nm versus 852 nm) Rb atoms are not affected by the Cs MOT laser. For the Cs MOT, in turn, the dipole trap is a small pertubation (energy shift of $\approx k_\mathrm{B} \times$100 $\mu$K $\approx \mathrm{h} \times$2 MHz at full intensity, see Tab. \ref{table_parameter}) causing a light shift which does not impede the operation of the MOT. 

\begin{figure}
    \resizebox{0.5\textwidth}{!}{
    \includegraphics{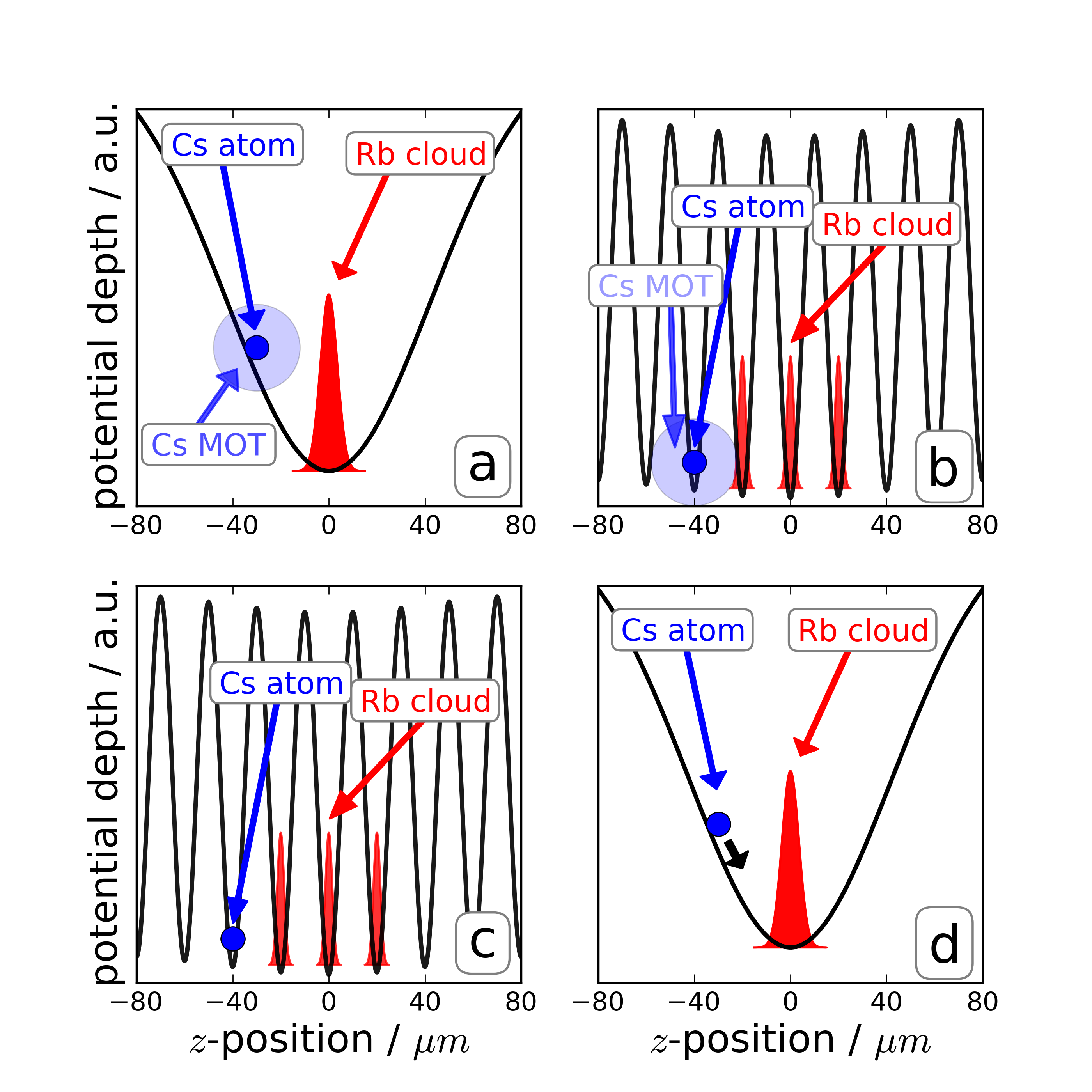}}
      \caption{Schematic sequence to insert a single Cs atom into a Rb gas (not to scale). \textbf{a)} An ultracold Rb gas (red) is prepared in the crossed dipole trap in the $|1,0\hspace{-.1cm}>$-state, with a high gradient MOT in close vicinity. Single Cs atoms are captured from the background gas. \textbf{b)} A lattice is adiabatically ramped up, confining both species in its optical potential. \textbf{c)} After switching off the Cs MOT both species are stored in the conserative lattice potential in separated lattice sites. \textbf{d)} The lattice is ramped down (to zero intensity) adiabatically, thereby transferring both species into the original crossed dipole trap.}
\label{sequence}
\end{figure}

The transfer of Rb into the magnetic-insensitive $|1,0\hspace{-.1cm}>$-state is completed before switching on the Cs MOT. With a homogeneous offset field of about 10 G applied, transitions between the two hyperfine states of Rb are driven. A first $\pi$-pulse at a frequency around 6.8 GHz transfers the Rb sample from the initial $|2,2\hspace{-.15cm}>$-state to the the $|1,1\hspace{-.15cm}>$-state, with a pulse duration of $\approx$ 19 $\mu$s, corresponding to a Rabi frequency of $\omega_{|2,2>\rightarrow|1,1>} \approx 2 \pi \times 26$ kHz. Accordingly, a following rf $\pi$-pulse around 7 MHz prepares the atoms in the $|1,0\hspace{-.1cm}>$-state. In this case, the pulse duration is $\approx$ 160 $\mu$s, corresponding to a Rabi frequency of $\omega_{|1,1>\rightarrow|2,0>} \approx 2 \pi \times 3.1$ kHz. The efficiency of this transfer is typically better than 95\%. Rb atoms remaining in a $m_F \neq 0$-state due to an imperfect pulse efficiency are removed from the trap by the gradient of the Cs MOT. 

Subsequently, the Cs MOT is switched on, with its center slightly offset from the Rb gas (in $z$-direction at $z \approx$ 30 $\mu$m, Fig. \ref{sequence}a). During the low gradient loading stage, where the geometrical size of the Cs MOT is large, the Cs density is low enough to avoid strong light-induced Rb-Cs collisions. This ensures that a sufficient number of Cs atoms can always be trapped, independent of the relative position of Cs MOT and Rb gas. In the high gradient stage, the density increases, and the loss rate can be employed for a precise relative positioning as explained below. The waist of the axial dipole trap is chosen to be $w_\mathrm{axial}=$100 $\mu$m, a few times larger then the diameter of the (high gradient) Cs MOT of about 30 $\mu$m. With the Cs MOT running, we transfer both species into the one-dimensional lattice (Fig. \ref{sequence}b). After switching off first the MOT laser and then the magnetic quadrupole field, both species are stored in purely optical potentials while being still separated (Fig. \ref{sequence}c). Then, Cs is optically pumped into the desired target state. Optionally, the internal state of Rb can as well be manipulated by rf/microwave radiation.

\begin{figure}
    \resizebox{0.5\textwidth}{!}{
    \includegraphics{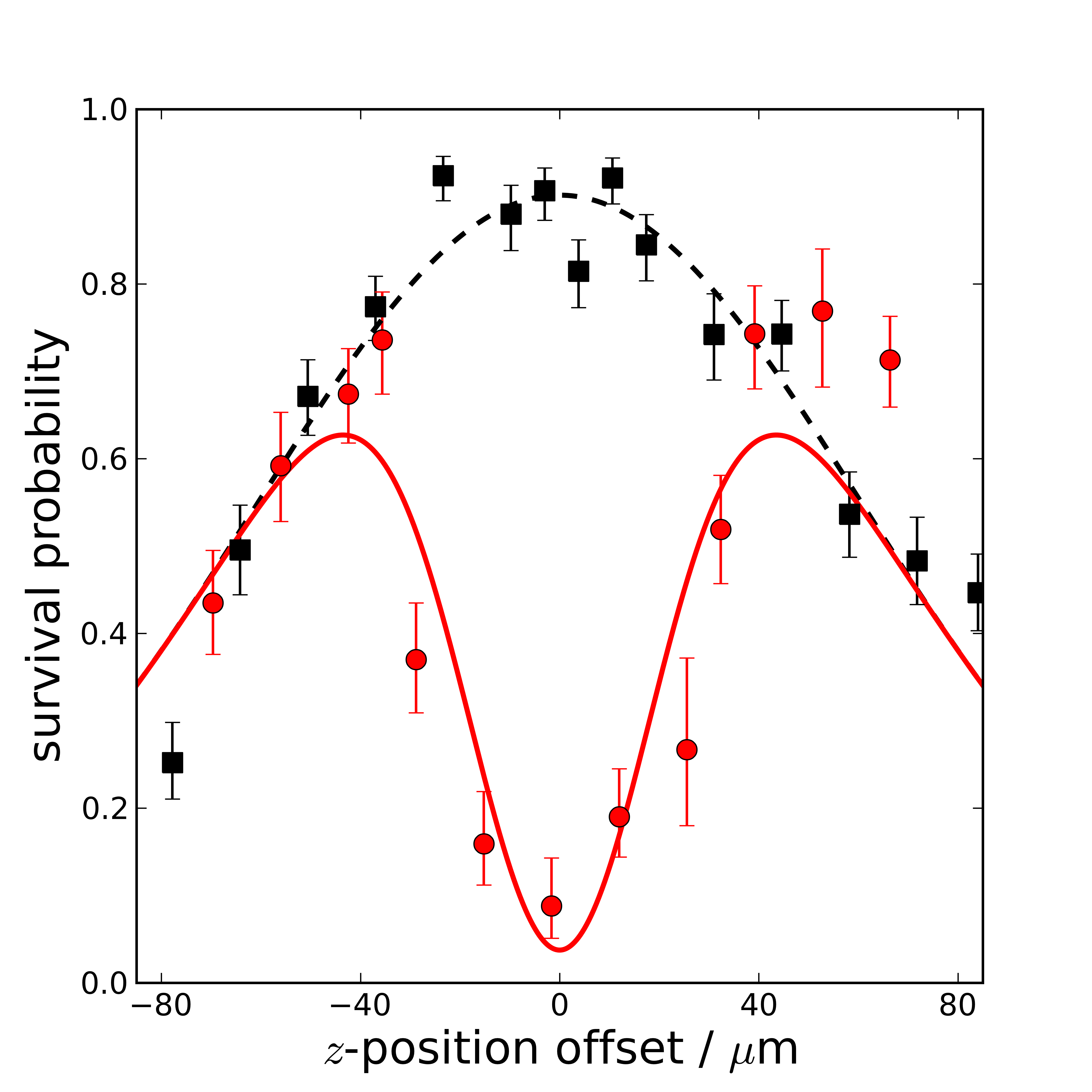}
    }
     \caption{Fine tuning the relative distance of Cs single atom MOT and Rb cloud using light-induced collisions. Black squares: Without Rb stored in the trap, the survival probability of Cs transferred to the crossed dipole trap is measured for different positions of the axial dipole trap in z-direction. Red circles: The same measurement with a Rb gas stored in the crossed dipole trap. }
\label{fig_ueberlapp}
\end{figure}

In the last step of the sequence, the lattice is ramped down adiabatically and the potential is transformed back into a running wave (Fig. \ref{sequence}d). Cs is cooled by this adiabatic expansion from a temperature of initially 27 $\mu$K to a temperature of $\approx$ 4 $\mu$K \cite{spethmann_nicolas_et_al._manuscript_2011}. At some lattice depth, the Cs atom leaves its lattice site and is trapped in the running wave dipole trap together with the Rb gas. From this point on, Rb and Cs interact in the conserative potential of the dipole trap. Here, however, we restrict the discussion to the fine positioning employing light-induced collisions of the two species. Ground state interactions between single Cs atoms and the ultracold Rb gas will be presented elsewhere \cite{spethmann_nicolas_et_al._manuscript_2011}.

\subsection{Light-induced collisions and fine positioning}

During the operation of the Cs MOT, the Cs atom scatters photons with a rate on the order of MHz. If in contact with the Rb cloud, the presence of near resonant light causes interspecies light-induced collisions \cite{weber_single_2010,weiner_experiments_1999}. These collisions provide a strong inelastic loss channel, causing Cs loss in a short time from the MOT if overlapped with Rb. In this regard a clear separation of the Cs MOT from the Rb cloud is favorable. To transfer the Cs atom into the running wave dipole trap, in contrast, the MOT should be as close as possible to the center of the trap, where the Rb is stored. A compromise between these two requirements allows to find an optimal position with a transfer efficiency of $\approx$ 80\%. 

In order to find this optimal position, we first determine the transfer efficiency of Cs atoms into the running wave trap without Rb present. By superimposing the spatial modes of radial dipole trap and the lattice beams we ensure a good alignment for the radial dipole trap. This leaves the $z$-direction as the only free parameter influencing the transfer efficiency, which can be adjusted by the axial dipole trap. In Fig. \ref{fig_ueberlapp} we measured this transfer efficiency as the survival probability of Cs in the running wave using the above mentioned sequence for different relative $z$-positions of the axial dipole trap without Rb present (black squares). We obtain a maximum survival probability after a storage time of one second of about 90 \%, where we attribute the slight reduction in survival probablity to transfer loss as well as collions with the background gas. For an intentional misalignment, the survival probability decreases: For a displacement of about 50 $\mu$m, the survival probability is reduced roughly by a factor of two. The data of this experiment was fitted with a Gaussian profile $p_\mathrm{{s}} \exp(-z^2/w_\mathrm{YAG}^2)$, where $p_\mathrm{{s}}$ is the amplitude of the survival probability and $w_\mathrm{YAG}$ the waist of the Gaussian. The fit shows a reasonable agreement. As expected, the obtained width $w_\mathrm{YAG}=86\mu$m roughly corresponds to the waist of the axial dipole trap of $100\mu$m (see Tab. \ref{table_parameter}).

In a second step, we store Rb in the trap, which drastically changes the situation. In a typical sequence, we cool Rb by evaporation in the dipole trap to a temperature of 100-200 nK, with Rb atom numbers of a few $10^4$. To decrease the size of the Rb gas and to provide a sufficient trap depth for the Cs atom, the trap is adiabatically compressed by ramping up the power of the dipole traps again. The temperature of the Rb gas rises due to this compression to several hundred nK, the size of the Rb cloud is $\approx$ (7,7,14) $\mu$m and the density is in the low 10$^{13}$/cm$^3$ range. Then the Cs MOT is switched on (Fig. \ref{sequence}a) and the lattice is ramped up (Fig. \ref{sequence}b). Here, any overlap between the two species will lead to loss due to light-induced collisions. Hence, upon varying the relative position between Cs MOT and Rb gas by changing the axial dipole trap position, the fraction of surviving Cs atoms will map out the effective interspecies overlap.

To prevent ground state inter-species collisions in the dark for this positioning measurement, we push Rb out off the trap immediately after loading into the lattice with a short resonant light pulse, directly after switching off the Cs MOT (Fig. \ref{sequence}c) and before ramping down the lattice (Fig. \ref{sequence}d). We have verified that this pulse does not affect the Cs atoms at all. This push out ensures that the two species are interacting via light-induced collisions only. Light-induced collisions during recapture of Cs are thus also avoided.

Assuming inelastic Cs-Rb two-body collisions as the dominant loss mechanism, the loss probability of Cs atoms is directly proportional to the density-density overlap of Rb and Cs. The density distributions for a thermal Rb cloud as used in this experiment as well as for the compressed Cs MOT can be described by a Gaussian with widths of $w_\mathrm{Rb}\approx$ 14 $\mu$m (Rb gas) and $w_\mathrm{Cs}\approx$ 30 $\mu$m (Cs MOT). To provide a model for this loss, a Gaussian $p_\mathrm{{loss}} \exp(-z^2/w_\mathrm{loss}^2)$ was substracted from the survival probability of Cs without Rb. The measurement presented in Fig. \ref{fig_ueberlapp} shows the experimental data and a fit with a function corresponding to this model. For distances of more then $\approx$ 40 $\mu$m between Cs MOT and the center of the trap, the survival probability is the same as without Rb present. For smaller distances, however, the survival probability is reduced. With the MOT in the center of the trap, essentially no Cs atom survives, indicating a matched overlap between Cs and Rb. 

The agreement between our model (red solid line in Fig. \ref{fig_ueberlapp}) and the data is reasonable. The width $w_\mathrm{loss}$ of the ''loss feature`` is given by the convolution of the density profiles of both species and thus expected to be $w_\mathrm{loss, theo}=\sqrt{w_\mathrm{Cs}^2+w_\mathrm{Rb}^2}\approx$ 33 $\mu$m, close to the value of $w_\mathrm{loss}=27$ $\mu$m obtained from the fit. However, there are significant deviations. The slope of the loss feature ($0<|z|\approx\pm$ 40 $\mu$m) appears to be steeper than expected from our simple model. A reason for this could be the onset of Cs-Rb-Rb three-body losses for higher Rb densites in the center of the trap. For our purpose, however, it is sufficient to experimentally determine the point of highest survival probability for Cs with Rb present, at $z\approx\pm$ 30 $\mu$m. At this point, the insertion of single Cs atoms into the Rb cloud has an efficiency of close to 80\%.

\section{Conclusions and outlook}
We have realized the insertion of single atoms into an ultracold gas with an efficiency close to 80\%. The ultracold gas and the single atom are cooled and trapped independently and then transferred into a common trap. Light-induced collisions are employed for fine positioning. By using a species-selective lattice, both species can be trapped in a purely optical trap despite their different temperatures and density regimes. This allows to manipulate the internal degree of freedom of both species before bringing the single atom into contact with the ultracold gas. Our setup allows to study the dynamics of single or few impurity atoms in a conserative potential. We have observed thermalization of single Cs atoms to the temperature of the Rb gas via elastic collisions \cite{spethmann_nicolas_et_al._manuscript_2011}. 

Due to the remaining selectivity of the trap at 1064 nm, it should be possible to cool a high density thermal Rb cloud to quantum degeneracy without loosing the Cs atom. Alternatively, one could start with a BEC and probe the dynamical insertion of the single atom into the BEC. This would allow the study of BECs that are doped with single atom impurities, opening the way to investigations of polaron-type physics \cite{cucchietti_strong-coupling_2006,kalas_interaction-induced_2006}. Another possibility would be the study of the coherence properties of the single atom immersed in the quantum gas \cite{ng_single-atom-aided_2008} and possible coherent cooling mechanisms of the single atom \cite{daley_single-atom_2004}.  

\textbf{Acknowledgements:}
We gratefully acknowledge support from the science ministry of North Rhine-Westfalia (NRW MWIFT) through an independent Junior Research Group. N.S. acknowledges support from Studienstiftung des Deutschen Volkes and N.S. and C.W. from the Bonn-Cologne Graduate School.


\begin{thebibliography}{}
\bibitem{bloch_many-body_2008}
I. Bloch, J. Dalibard, W. Zwerger, Reviews of Modern Physics \textbf{80,} 885 (2008)
\bibitem{cucchietti_strong-coupling_2006}
F.M. Cucchietti, E. Timmermans, Physical Review Letters \textbf{96,} 210401 (2006)
\bibitem{kalas_interaction-induced_2006}
R.M. Kalas, D. Blume, Physical Review A \textbf{73,} 043608 (2006)
\bibitem{griessner_dark-state_2006}
A. Griessner, A.J. Daley, S.R. Clark, D. Jaksch, P. Zoller, Physical Review Letters \textbf{97,} 220403 (2006)
\bibitem{bakr_quantum_2009}
W.S. Bakr, J.I. Gillen, A. Peng, S. F\"olling, M. Greiner, Nature \textbf{462,} 74--77 (2009)
\bibitem{sherson_single-atom-resolved_2010}
J.F. Sherson, C. Weitenberg, M. Endres, M. Cheneau, I. Bloch, S. Kuhr, Nature \textbf{467,} 68--72 (2010)
\bibitem{gericke_high-resolution_2008}
T. Gericke, and P. Wurtz, D. Reitz, T. Langen, H. Ott, Nat Phys \textbf{4,} 949--953 (2008)
\bibitem{zipkes_trapped_2010}
C. Zipkes, S. Palzer, C. Sias, M. K\"ohl, Nature \textbf{464,} 388--391 (2010)
\bibitem{schmid_dynamics_2010}
S. Schmid, A. H\"arter, J. Hecker Denschlag, Physical Review Letters \textbf{105,} 133202 (2010)
\bibitem{foerster_microwave_2009}
L. F\"orster, M. Karski, J.M. Choi, A. Steffen, W. Alt, D. Meschede, A. Widera, E. Montano, J.H. Lee, W. Rakreungdet, P.S. Jessen, Physical Review Letters \textbf{103,} 233001 (2009)
\bibitem{PhysRevLett.102.053001}
M. Karski, L. F\"orster, J.M. Choi, W. Alt, A. Widera, D. Meschede, Physical Review Letters \textbf{102,} 053001 (2009)
\bibitem{esslinger_bose-einstein_1998}
T. Esslinger, I. Bloch, T.W. H\"ansch, Physical Review A \textbf{58,} R2664 (1998)
\bibitem{haas_species-selective_2007}
M. Haas, V. Leung, D. Frese, D. Haubrich, S. John, C. Weber, A. Rauschenbeutel, D. Meschede, New Journal of Physics \textbf{9,} 147--147 (2007)
\bibitem{weber_single_2010}
C. Weber, S. John, N. Spethmann, D. Meschede, A. Widera, Physical Review A \textbf{82,} 042722 (2010)
\bibitem{ueberholz_b._cold_2002}
B. Ueberholz, S. Kuhr, D. Frese, V. Gomer, D. Meschede, Journal of Physics B: Atomic, Molecular and Optical Physics \textbf{35,} 4899--4914 (2002)
\bibitem{alt_single_2003}
W. Alt, D. Schrader, S. Kuhr, M. M\"uller, V. Gomer, D. Meschede, Physical Review A \textbf{67,} 033403 (2003)
\bibitem{tuchendler_energy_2008}
C. Tuchendler, A.M. Lance, A. Browaeys, Y.R.P. Sortais, P. Grangier, Physical Review A \textbf{78,} 033425 (2008)
\bibitem{leblanc_species-specific_2007}
L.J. {LeBlanc}, J.H. Thywissen, Physical Review A \textbf{75,} 053612 (2007)
\bibitem{spethmann_nicolas_et_al._manuscript_2011}
N. Spethmann et al., manuscript in preparation
\bibitem{weiner_experiments_1999}
J. Weiner, V.S. Bagnato, S. Zilio, P. S. Julienne, Reviews of Modern Physics \textbf{71,} 1 (1999)
\bibitem{ng_single-atom-aided_2008}
H.T. Ng, S. Bose, Physical Review A \textbf{78,} 023610 (2008)
\bibitem{daley_single-atom_2004}
A.J. Daley, P.O. Fedichev, P. Zoller, Physical Review A \textbf{69,} 022306 (2004)

\end{thebibliography}

\end{document}